\documentstyle[aps,prb]{revtex}
\begin{document}
\draft
\title{Dilution Effects on Ordering in the ${S=1/2}$ Heisenberg
       Antiferromagnet \protect \\ on the Square Lattice}
\author{J. Behre}
\address{Institut f\"ur Theoretische Physik, Universit\"at
  Hannover, Appelstr.~2, 3000 Hannover 1, Germany
    \\  email (internet): behre@kastor.itp.uni-hannover.de
        }
\author{S. Miyashita}
\address{Graduate School of Human \& Environmental Studies,
  Kyoto University, Kyoto, Japan
    \\  email (Bitnet): miya@JPNYITP
        }

\date{\today}
\maketitle
\begin{abstract}
 The influence of dilution with non--magnetic impurities on the order
 in the ground state of the $S=1/2$ Heisenberg antiferromagnet is
 investigated by Quantum Monte Carlo simulations. Data of the spin
 correlation functions and thermodynamic properties for system sizes up
 to $L\times L=16\times16$ and for impurity concentrations up to
 $\delta=37.5\%$ are presented. In the low doping regime the
 correlation length shows similiar dependence on temperature as in the
 pure case. The staggered magnetization $\langle N_z^2(\delta)\rangle$
 in the ground state which is the order parameter of the present model
 is estimated by extrapolating data at low temperatures. Impurity
 concentration dependence of $\langle N_z^2(\delta)\rangle$ is
 presented. $\langle N_z^2(\delta)\rangle$ decreases monotonically
 with impurity concentration and becomes very small around
 $\delta_c \sim 0.35$ which is much smaller than the classical
 percolation threshold.
\end{abstract}


\section{Introduction} \label{Introduction}

 In this article we show new results on the spin correlation functions
 and the long range order of the diluted Heisenberg antiferromagnet on
 the square lattice with $S=1/2$. Originally this work was intended to
 understand the effects created upon doping of the high temperature
 superconductor materials. It is now generally believed that the pure
 antiferromagnet with $S=1/2$ on the square lattice has long range
 order~\cite{reg,chn}. But the value of the order parameter is much
 reduced against its classical value due to quantum fluctuations.
 The interesting point now is the influence of inhomogeneities on the
 competition between order and quantum fluctuations.

 For our purpose it is necessary to study large lattices and these
 systems are too large to use exact diagonalization. Therefore we used
 the Quantum Monte Carlo method (QMC). Because we are working only
 with static impurities, the minus sign problem does not suffer the
 present QMC. And our results might be applicable to experiments with
 materials like
 \mbox{$La_2 Cu_{1-\delta} Mg_\delta O_4$} and
 \mbox{$La_2 Cu_{1-\delta} Zn_\delta O_4$} \cite{cheong}.
 In these materials magnetic copper atoms are substituted by
 non--magnetic zinc or magnesium atoms, so there are static
 non--magnetic site impurities, which are the same type of impurities
 we are considering in our simulations.

 In the present paper we report the temperature and impurity
 concentration dependence of the spin correlation functions and also
 the long range order in the ground state.

 We have studied the spin correlation functions for various
 concentrations in a wide temperature region including lower
 temperatures than the previous Manousakis' work~\cite{manous3}. We
 analyze the concentration dependence of the correlation length in the
 form
 \begin{equation}
    \xi(T,\delta)=C_\xi(\delta) e^{\frac{2\pi\rho_S(\delta)}{T}} \;.
 \end{equation}
 The correlation data fit this behavior well. And the estimated spin
 stiffness $\rho_S(\delta)$ is monotonically decreasing with impurity
 concentration.

 In the classical analog of the model, the long range order exists as
 long as the lattice itself percolates. Namely, the long range order
 vanishes at an impurity concentration corresponding to the percolation
 threshold, $\delta_c=0.41$~\cite{essam}. In the present paper, the
 impurity concentration dependence of the long range order is reported.
 We would like to point out that the amount of the long range order
 becomes very small for $\delta \sim 0.35$, which is much smaller than
 $\delta_c=0.41$. The concentration dependence of percolating lattice
 sites has a dependence $(\delta-\delta_c)^{\beta}$ with $\beta < 1$
 and therefore it has a convex shape. On the other hand, the obtained
 dependence of the long range order for the present model has a concave
 shape. Furthermore, it is suggested that the critical value of the
 concentration is also reduced in the quantum case.

 Our method is described in chapter~\ref{Method}. The results for the
 spin--spin correlation functions are shown in
 chapter~\ref{Correlation} and those for the long range order in
 chapter~\ref{Order}. The results are summarized and discussed in
 chapter~\ref{Discussion}.

\section{Method} \label{Method}

 The model including the inhomogeneities used here can be described by
 the Hamiltonian
 \begin{equation}
    {\cal H} =J\sum_{<i,j>} \varepsilon_i \varepsilon_j
     \vec{S_i} \cdot \vec{S_j}\;,
 \end{equation}
 where $\varepsilon_i =0$ for holes otherwise $\varepsilon_i=1$ (for spins),
 $\vec{S_i}$  are the $S=1/2$ spin operators
 ($\vec{S_i}=\frac{1}{2}\vec{\sigma_i}$),
 and the sum runs over nearest neighbour pairs $i$ and $j$, $J$ is the
 antiferromagnetic coupling constant. The positions of the holes were
 chosen randomly but fixed for one simulation, so they are static
 non--magnetic impurities. The concentration of impurities is defined
 by $\delta$,
 \begin{equation}
    \delta =1-\sum_i \varepsilon_i/L^2 \;,
 \end{equation}
 where $L$ is the linear system size of the square lattice and
 $N=L\times L$ is the number of lattice sites.

 For the simulation we used the Quantum Monte Carlo method based on the
 Suzuki--Trotter formula and checker board
 decomposition~\cite{suzuki,miya1,reg}.
 Periodic boundary conditions were applied to the $x$--, $y$-- and Trotter
 directions. In addition to standard local flip types we used special flip
 types in the neighbourhood of the impurities, local loop flips around
 small clusters of holes or single holes and non straight global flips
 in the $x$--$y$ plane. Details of these new flip types are described
 in~\cite{behre3}. These new flip types are necessary for the ergodicity
 of the algorithm.

 We study system sizes $L\times L$ with $L=4,8,12,16\,$ for impurity
 concentrations up to $\delta=37.5\%$ corresponding to $6,24,36,96\,$
 holes. We perform simulations at various temperatures. The lowest one
 is $T=0.05J$. For this low temperature it is necessary to go up to
 $9\cdot 10^5$ Monte Carlo steps and Trotter numbers $m=32$. We perform
 simulations for each sample with 3 or 4 different Trotter numbers $m$
 and then extrapolate for $m$ to infinity by using a linear function in
 $(\frac{1}{m^2})$. An extrapolation up to quadratic order is used as a
 test for the error of the extrapolation. From these data the ground
 state properties are obtained by extrapolations.

 The data for $L=4$ are available from exact diagonalization. From
 these data we estimate the low temperature properties by summing up
 several low lying levels with the canonical weight. They are in good
 agreement with the data obtained by QMC. Larger systems are simulated
 by QMC. For each system size and hole concentration we average the
 data over five different samples with hole positions chosen randomly.
 Between them we found only small variation. Averaging over more than
 five samples is too much computer time consuming.

 To study such a big parameter regime it is necessary to make an
 optimized and fully vectorized program. Locations of the new type
 flips for a given hole configuration are listed up and stored before
 the simulation. All transition probabilities for each fliptype, which
 are given by ratios of Boltzmann factors between possible local spin
 configurations, are also stored in tables in the beginning of the run.
 During the simulation the computer only has to compare the transition
 probability for a flip with a random number. If the flip is accepted,
 the configuration is replaced by the flipped configuration which is
 also stored in another table.

 After one sweep, namely after each flip type is tested over
 the whole lattice, physical quantities are calculated. The data of
 10000 sweeps are combined to one bin. The samples for each bin seem to
 be statistically independent with Gaussian distribution, so they can
 be used to calculate statistical errors. These data are extrapolated
 for $m$ to infinity as mentioned above and then averaged over
 different configurations with the same system size and hole
 concentration. Averaging and extrapolation is done using a weighting
 procedure, each data point is weighted by the reciprocal of the square
 of its error. Final error bars are estimated only by this procedure.
 The further treatment of the correlation data are
 described in chapter~\ref{Correlation} and the sublattice
 magnetization in chapter~\ref{Order}.

\section{Concentration Dependence of the Spin Correlation Functions}
 \label{Correlation}

 One interesting property of the system is how the spin correlation function
 depends on the distance between two spins
 \begin{equation}
    c(r) =(-1)^r \frac{4}{L^2}\sum_i \langle S_i^z S_{i+r}^z \rangle \;,
    \label{scr}
 \end{equation}
 where we only change the positions along the $x$-- or $y$--direction.

 In Fig.~(\ref{cr}) the spin correlation functions for the system size
 $L=16$ and temperature $T=0.1J$ is shown for different hole
 concentrations. The correlation functions decrease with distance,
 although the data are symmetric at $r=8$ because of the periodic
 boundary conditions. They also monotonically decrease with hole
 concentration for each distance. In references
 \cite{behre2,miya2,behre3} the enhancements of short range correlation
 function near impurities are reported. But in the normalization of
 Eq.~(\ref{scr}) the enhancements by nearest neighbour correlations
 near impurities are compensated by decreasement of the number of pairs
 of spins.

 In order to study the temperature dependence of the correlation
 length, we have to fit our data to the expected correlation function.
 There is no agreement in the literature about the right form of this
 function as can be seen in the discussion of~\cite{barnes1}.
 Ding and Makivi\'c~\cite{ding} used the form
 \begin{equation}
    c(r)=Ar^{-\lambda}e^{-\frac{r}{\xi}} \label{corr} \;,
 \end{equation}
 where they used both of $A$ and $\lambda$ as the fitting parameters
 and found $\lambda\approx 0.4$. However spin wave
 calculations~\cite{takahashi} and the Schwinger boson
 method~\cite{arovas} give $\lambda=1$, the renormalization group
 approach~\cite{chn} gives $\lambda=0.5$ and some other
 authors~\cite{gomez,manous1,manous3} do not use the algebraic decaying
 part corresponding to $\lambda=0$.

 If we generalize the temperature dependence of the correlation length
 proposed by Chakravarty et.~al.~\cite{chn} to the case with
 impurities, we have~\cite{manous3}
 \begin{equation}
    \xi(T,\delta)=C_\xi(\delta)e^{\frac{2\pi\rho_S(\delta)}{T}} \;.
    \label{crlen}
 \end{equation}
 The prefactor $C_\xi$ and the spin stiffness $\rho_S$ now depend on
 the hole concentration $\delta$ and we want to test the validity of the
 form and specify the parameter regime where the form is valid.

 We tried the least square fits of the data of the correlation
 function to the form Eq.~(\ref{corr}) with different values of
 $\lambda$. In Fig.~(\ref{cr}) the fitting curves using $\lambda=1/2$
 are shown. The fits look excellent and the errors for the fitted
 correlation lengths are very small (if we exclude the nearest
 neighbour spin pairs for the fit). If we use $\lambda=1$ we get
 nearly the same results if the fitted correlation length is smaller
 than the system size. We find no significant difference for these two
 values of $\lambda$. But if we put $\lambda=0$ we find bigger
 deviations of the fitting functions from the data points, and the
 fitting errors are much enhanced. In our analysis, there are too
 small number of data points to use $\lambda$ also as a fitting
 parameter. If we would use the two parameters, $\lambda$ and $\xi$, as
 the fitting parameter, $\lambda$ varies largely to fit the shapes. In
 particular, for the cases where we found a long correlation length by
 using a fixed value of $\lambda$, the two parameter fitting concludes
 an algebraic decay without exponential part, namely $\lambda \ne 0$
 and $\xi\simeq 0$.

 So we estimated the correlation lengths by fitting using
 $\lambda=1/2$. Their logarithmic values are plotted in Fig.~(\ref{xi})
 over $1/T$ for the different hole concentrations. Only the correlation
 lengths which are smaller than the system size are plotted. For these
 data the size dependence is small. Data appear in straight lines in
 this graph. Thus the temperature dependence is consistence with
 Eq.~(\ref{crlen}) and we can estimate the spin stiffness $\rho_S$ from
 the slopes. In this graph error bars are not drawn because some of
 them are overlapping and they are rather confusing for the eye. For
 temperatures higher than $T=0.5J$ the error bars are nearly like the
 point size. At $T=0.25J$, the error bars for low concentrations, where
 the correlation length becomes comparable to the system size, increase
 up to 10\%. In the estimation of $\rho_S$, this fact is taken into
 account by using a weighted fitting procedure. The fits are
 acceptable, but for high concentrations other fit functions cannot be
 excluded.

 The concentration dependence of the spin stiffness is given in
 Fig.~(\ref{rhos}) and the estimated prefactor $C_\xi (\delta)$ in
 Fig.~(\ref{cxi}). For the pure system $(\delta=0)$ we find
 \begin{eqnarray}
    \rho_S &=& 0.1734\pm 0.0003J \nonumber\\
    C_\xi  &=& 0.393 \pm 0.008\;.
 \end{eqnarray}
 The shown errors are those which come only from our weighted fitting
 procedure and are definitely too small. The values are in the range
 of other authors as discussed in \cite{barnes1}. The prefactor
 slightly increases with concentration up to $\delta=0.25$. The spin
 stiffness decreases by more than a factor of two in this regime and
 saturates afterwards. This behavior does not agree with the scaling
 properties proposed by Manousakis~\cite{manous3}, who used a smaller
 parameter regime.

 Recently Yanagisawa~\cite{yanagisawa} has predicted
 making use of the non--linear$\sigma$--model with impurities that
 the exponential dependence on the temperature
 Eq.~(\ref{crlen}) changes to a linear dependence $1/T$
 at a very small impurity concentration and that
 $\xi$ is a temperature independent constant for $\delta$ larger than
 some critical concentration $\delta_c$ where the system is not ordered.
 The present data do not quantitatively agree with the work
 although we cannot rule out a transition to the linear dependence.

\section{Concentration Dependence of the Long Range Order}
\label{Order}

 In this chapter the data for the long range order in the ground state
 are studied. The order parameter of the system is the sublattice
 magnetization
 \begin{equation}
    \langle N_z^2\rangle=\frac{1}{N^2} \biggl\langle \biggl(
    \sum_{i\epsilon \cal A} \sigma_i^z - \sum_{j\epsilon\cal B}
    \sigma_j^z \biggr)^2 \biggr\rangle\;,
 \end{equation}
 where $\cal A$ and $\cal B$ denote the two different sublattices. In
 Fig.~(\ref{nz}a) the sublattice magnetization is plotted against the
 temperature for different hole concentrations and for the system size
 $L=8$. Fig.~(\ref{nz}b) shows the data for $L=12$ and Fig.~(\ref{nz}c)
 for $L=16$. For $L=8$ there is no big variation for the low temperatures
 and the extrapolation to $T=0$ is smooth, which means that the system
 is near the ground state at the low temperatures. But for $L=12$ and
 $L=16$ the errorbars are enlarged at low temperatures, although
 we could estimate a general tendency of the concentration dependence.
 In order to obtain these data, up to $9\cdot 10^5$ MCS have been
 performed for lattices with several Trotter numbers. Still the
 errorbars are rather large. This shows the general difficulty to
 obtain low temperature data with QMC.

 We regard the data at the lowest temperature as the ones in the
 ground state and extrapolate these data to the thermodynamic limit,
 namely $L$ to infinity. The extrapolation is done using a polynomial
 in $1/L$. This is the same extrapolation scheme as in~\cite{reg},
 where it is assumed that the spin correlation function decays by a
 power law $c(r) \sim {1\over r}$ as is indicated by the spin wave
 theory. In Fig.~(\ref{nzl}) the data are plotted against $1/L$ for
 some concentrations. The size dependences are smoothly extrapolated as
 well as the pure case. Two different extrapolations are shown, a
 linear one and one with a polynomial up to the quadratic order. The
 extrapolated value for $\delta=0.375$ is less than zero. That means
 the extrapolation scheme is no longer applicable here and the order
 parameter would already vanish.

 To compare our sublattice magnetization with the spontaneous
 magnetization $m^\dagger =\langle S_i^z\rangle$ calculated in the spin wave
 theory, one has to consider the rotational invariance in spin space
 and the difference between $S$ and $\sigma$. In the pure case
 ($\delta=0$) we find
 \begin{equation}
    m^\dagger =\sqrt{\frac{3}{4}\langle N_z^2\rangle}= 0.315\pm 0.04\;.
 \end{equation}
 There are many results of this quantity as are seen
 in review articles~\cite{barnes1,manous2}, e.~g.\
 Anderson~\cite{anderson52} found in spin wave calculations
 $m^\dagger =0.303$ and Reger and Young~\cite{reg} found
 $m^\dagger =0.30\pm 0.02$ with QMC. Our result is a little bit
 higher.

 In Fig.~(\ref{nzd}), the linear extrapolated data and the data at
 $T=0.1$ for various sizes are plotted against the concentration. The
 concentration dependences for each size are nearly described by a
 straight line. This suggests that there is only a small cooperative
 effect of the holes for the global order. The local effect found in
 \cite{behre3} seems to be averaged out. The error bars are large in
 the middle concentration regime. The deviation of data in different
 samples is also large here. This observation implies that the order
 parameter depends more on the special arrangement of the holes for
 these concentrations. Here we see the extrapolated long rage order
 decreases with the concentration. It should be noted that the
 concentration dependence has a concave shape in contrast to a convex
 shape of the concentration dependence of the number of the percolating
 cluster which is equivalent to the long range order in the classical
 model. This suggests that quantum fluctuations have a stronger effect
 at higher concentrations. In the theory of the non--linear--$\sigma$
 model, it is suggested that there is a critical value for the quantum
 fluctuations. If the system has smaller quantum fluctuations than
 the critical value, the quantum fluctuations are irrelevant and the
 classical picture of ordering is realized. For example, the pure case
 belongs to this case and also cases with low concentrations obviously
 belong to this case. If the concentration becomes large, however,
 quantum fluctuations might exceed the critical value. Then the system
 shows a kind of a quantum disordered state where the long range order
 disappears although the lattice itself percolates. This critical value
 may correspond to $\delta_c$ in the reference~\cite{yanagisawa} and
 then his conclusion qualitatively agree with the present results.

 Although the fluctuation is rather large for definite conclusion, we
 would estimate the lowest concentration where the order parameter
 becomes zero in Fig.~(\ref{nzd}). We interpolate the data points and
 find roughly
 \begin{equation}
    \delta_c=0.345\pm 0.015\;.
 \end{equation}
 The error is only interpolated from the errorbars of the extrapolated
 data points and might be too small. In order to obtain a better
 value of the critical concentration, it is necessary to simulate
 larger systems at lower temperatures near our critical concentration
 and to average over more different hole configurations. But at present
 this would exceed the ability of standard computers. However the above
 estimation implies that the critical concentration is lower than the
 classical percolation threshold $\delta_c=0.41$~\cite{essam}.  This
 point should be studied more in the future.

 There are only very few studies on the doped case.
 Pimentel and Orbach~\cite{pimentel} find in spin wave calculations to
 the $t$--$J$ model $\delta_c=0.32$, which they interpret as an upper
 bound for the sublattice magnetization. But the order in their model
 is destroyed by the mobility of the impurities.

\section{Summary and Discussion}  \label{Discussion}

 We have extended the usual QMC algorithm to treat inhomogeneities in a
 lattice. Our new flip types enabled us to simulate diluted Heisenberg
 antiferromagnets on the square lattice. Results on the correlation
 length and the long range order have been presented.

 From the spin correlation function, temperature and concentration
 dependences of the correlation lengths are investigated. The
 correlation lengths show an exponential dependence on $1/T$ over a
 wide concentration regime. The spin stiffness decreases more than 50\%
 when the impurity concentration increases from zero to 0.25. It
 saturates around $\delta = 0.25$ and it seems to stay finite even at
 concentrations where the sublattice magnetization vanishes. But data
 for high concentrations have a rather large deviation and we do not
 exclude that the temperature dependence of the correlation length
 changes from the exponential form to a linear form at high
 concentrations.

 The sublattice magnetization at $T=0$ is estimated by extrapolation
 of data with finite $L$ to infinity. For the pure case we get the
 same results of other authors. The sublattice magnetization of doped
 systems shows a similar size dependence as the pure case. We find a
 monotonic decrease of the order parameter with hole concentration.
 From the global shape of the concentration dependence, we conclude
 that the quantum fluctuations have a stronger effect at the high
 concentration region. Although the error bar is large, the critical
 concentration where the system becomes disordered is estimated as
 $\delta_c=0.345$. This value is smaller than the classical
 percolation threshold. It is suggested that quantum fluctuation causes
 a quantum disordered state in the ground state.

 It is difficult to compare directly the present data with the
 experimental data for doped \mbox{$La_2 Cu O_4$} because of the
 following two reasons. First, in the two-dimensional Heisenberg model
 the Mermin--Wagner theorem~\cite{mermin} forbids breaking of a
 continuous symmetry in two--dimensional models at finite temperatures,
 while in the materials there is a small interplane coupling which
 creates a non zero N\'eel--temperature. Second, in the materials,
 doping creates mobile holes which cause different effects from the
 static holes. But recently Cheong et.~al.~\cite{cheong} investigated
 \mbox{$La_2 Cu_{1-\delta} Mg_\delta O_4$} and
 \mbox{$La_2 Cu_{1-\delta} Zn_\delta O_4$}, where doping leads to
 non--magnetic static impurities, which is closer to the present model
 although they are still itinerant systems. They find that the
 N\'eel--temperature goes to zero for a concentration $\delta_c=0.22$
 and that the critical concentration is clearly different from
 materials which can be described by a model with $S=5/2$ or by an
 Ising--model. So there is also a hint for a reduction of the classical
 percolation threshold from experiments.

 We hope that the concentration dependence of the long range order
 which is presented in this paper is found experimentally in some real
 materials and furthermore that the problem, whether the $\delta_c$ and
 the percolation threshold is different or not, is solved in future.
 This seems difficult for numerical methods at the present situation.
 We also hope for the appearance of very high performance special
 computers which may solve the problem.

 The authors would like to thank Professor Mikeska for his continuous
 encouragement. The numerical simulations for the present study was
 very extensive. We spent more than 1500 CPU--hours on the FUJITSU S400/40
 (5 Gflops peak performance) of the Regionales Rechenzentrum
 Niedersachsen and several thousand hours on IBM RS6000 workstations.
 The present study has been partially supported by a Grant-in-Aid for
 scientific Research on Priority Area from Ministry of Education,
 Science and Culture of Japan.

  \begin{figure}
  \caption{The concentration dependence of the correlation function at
           $T=0.1$.}
  \label{cr}
  \end{figure}

  \begin{figure}
  \caption{The temperature dependence of the correlation length.}
  \label{xi}
  \end{figure}

  \begin{figure}
  \caption{The concentration dependence of the correlation parameter $\rho_S$.}
  \label{rhos}
  \end{figure}

  \begin{figure}
  \caption{The concentration dependence of the correlation parameter $C_\xi$.}
  \label{cxi}
  \end{figure}

  \begin{figure}
  \caption{Temperature-dependence of $\langle N_z^2 \rangle$ for (a) $L=8$
           (b) $L=12$ and (c) $L=16$.}
  \label{nz}
  \end{figure}

  \begin{figure}
  \caption{Size dependences of $\langle N_z^2 \rangle$}
  \label{nzl}
  \end{figure}

  \begin{figure}
  \caption{Concentration-dependence of $\langle N_z^2 \rangle$ for
           different system sizes and extrapolated values.}
  \label{nzd}
  \end{figure}

\end{document}